\documentstyle{mn}
\input epsf

\newif\ifAMStwofonts

\def\vlos{v_{\rm los}}
\def\half{\textstyle{1\over2}}

\ifoldfss
  \ifCUPmtlplainloaded \else
    \NewTextAlphabet{textbfit} {cmbxti10} {}
    \NewTextAlphabet{textbfss} {cmssbx10} {}
    \NewMathAlphabet{mathbfit} {cmbxti10} {} 
    \NewMathAlphabet{mathbfss} {cmssbx10} {} 
  \fi
  \ifAMStwofonts
    \ifCUPmtlplainloaded \else
      \NewSymbolFont{upmath} {eurm10}
      \NewSymbolFont{AMSa} {msam10}
      \NewMathSymbol{\upi}     {0}{upmath}{19}
      \NewMathSymbol{\umu}     {0}{upmath}{16}
      \NewMathSymbol{\upartial}{0}{upmath}{40}
      \NewMathSymbol{\leqslant}{3}{AMSa}{36}
      \NewMathSymbol{\geqslant}{3}{AMSa}{3E}

    \fi
  \fi
\fi 

\ifnfssone
  \newmathalphabet{\mathit}
  \addtoversion{normal}{\mathit}{cmr}{m}{it}
  \addtoversion{bold}{\mathit}{cmr}{bx}{it}
  \newmathalphabet{\mathbfit} 
  \addtoversion{normal}{\mathbfit}{cmr}{bx}{it}
  \addtoversion{bold}{\mathbfit}{cmr}{bx}{it}
  \newmathalphabet{\mathbfss} 
  \addtoversion{normal}{\mathbfss}{cmss}{bx}{n}
  \addtoversion{bold}{\mathbfss}{cmss}{bx}{n}
  \ifAMStwofonts
    \ifCUPmtlplainloaded \else
      \UseAMStwoboldmath
      \makeatletter
      \new@mathgroup\upmath@group
      \define@mathgroup\mv@normal\upmath@group{eur}{m}{n}
      \define@mathgroup\mv@bold\upmath@group{eur}{b}{n}
      \edef\UPM{\hexnumber\upmath@group}
      \new@mathgroup\amsa@group
      \define@mathgroup\mv@normal\amsa@group{msa}{m}{n}
      \define@mathgroup\mv@bold\amsa@group{msa}{m}{n}
      \edef\AMSa{\hexnumber\amsa@group}
      \makeatother
      \mathchardef\upi="0\UPM19
      \mathchardef\umu="0\UPM16
      \mathchardef\upartial="0\UPM40
      \mathchardef\leqslant="3\AMSa36
      \mathchardef\geqslant="3\AMSa3E
    \fi
  \fi
\fi 

\ifnfsstwo
  \DeclareMathAlphabet{\mathbfit}{OT1}{cmr}{bx}{it}
  \SetMathAlphabet\mathbfit{bold}{OT1}{cmr}{bx}{it}
  \DeclareMathAlphabet{\mathbfss}{OT1}{cmss}{bx}{n}
  \SetMathAlphabet\mathbfss{bold}{OT1}{cmss}{bx}{n}
  \ifAMStwofonts
    \ifCUPmtlplainloaded \else
      \DeclareSymbolFont{UPM}{U}{eur}{m}{n}
      \SetSymbolFont{UPM}{bold}{U}{eur}{b}{n}
      \DeclareSymbolFont{AMSa}{U}{msa}{m}{n}
      \DeclareMathSymbol{\upi}{0}{UPM}{"19}
      \DeclareMathSymbol{\umu}{0}{UPM}{"16}
      \DeclareMathSymbol{\upartial}{0}{UPM}{"40}
      \DeclareMathSymbol{\leqslant}{3}{AMSa}{"36}
      \DeclareMathSymbol{\geqslant}{3}{AMSa}{"3E}
    \fi
  \fi
\fi 

\ifCUPmtlplainloaded \else
  \ifAMStwofonts \else 
    \def\upi{\pi}
    \def\umu{\mu}
    \def\upartial{\partial}
  \fi
\fi

\title[Shell kinematics]
   {Measuring galaxy potentials using shell kinematics}
\author[M.R. Merrifield and K. Kuijken]
       {Michael R. Merrifield$^1$ and Konrad Kuijken$^{2,3}$\\
        ${}^1$Dept.~of Physics and Astronomy, 
	University of Southampton, S017 1BJ\\
        ${}^2$Kapteyn Instituut, P.O. Box 800, 9700 AV, Groningen, 
Netherlands\\
${}^3$Visiting Scientist, 
Dept.~of Theoretical Physics, University of the Basque Country, Lejona, Spain
}
\date{\today}
\pubyear{1997}

\begin{document}

\maketitle

\begin{abstract}

We show that the kinematics of the shells seen around some elliptical
galaxies provide a new, independent means for measuring the
gravitational potentials of elliptical galaxies out to large radii.  A
numerical simulation of a set of shells formed in the merger between an
elliptical and a smaller galaxy reveals that the shells have a
characteristic observable kinematic structure, with the maximum
line-of-sight velocity increasing linearly as one moves inward from a
shell edge.  A simple analytic calculation shows that this structure
provides a direct measure of the gradient of the gravitational
potential at the shell radius.  In order to extract this information
from attainable data, we have also derived the complete distribution
of line-of-sight velocities for material within a shell; comparing the
observed spectra of a shell to a stellar template convolved with this
distribution will enable us to measure the gradient of the potential
at this radius.  Repeating the analysis for a whole series of nested
shells in a galaxy allows the complete form of the gravitational
potential as a function of radius to be mapped out.  The requisite
observations lie within reach of the up-coming generation of large
telescopes.

\end{abstract}

\section{Introduction}

The gravitational potentials of elliptical galaxies have proved
difficult to derive, especially at large radii.  In disk galaxies,
both stars and gas have relatively simple orbit structures dominated
by nearly circular orbits, and it is therefore straightforward to
deduce the underlying gravitational potential (at least in the disk
plane) from the observed kinematics in these systems.  However, the
corresponding orbital structure in elliptical galaxies is much more
complicated, making the derivation of the gravitational potential from
observed kinematics in such a system far from simple.  In fact, basic
kinematic observations of projected density and line-of-sight velocity
dispersion are not sufficient to solve unambiguously for both the
functional form of the gravitational potential and the distribution of
stellar orbits in elliptical galaxies (Binney \& Mamon 1982).  This
ambiguity has recently been partially resolved by using the extra
kinematic information that can be derived from the shapes of line
profiles in high quality spectral data (e.g. Gerhard 1993, Carollo et
al.\ 1995). However, these results are restricted to the inner few
effective radii of the stellar light; it is hard to extend these
methods to much larger radii because of the low surface brightnesses
of galaxies further out.  Independent mass measurements using
gravitational lensing (e.g. Kochanek 1995) are restricted to still smaller
radii. The only technique that reaches to large radii comes from
mapping the distribution of hot gas around some ellipticals (Buote \&
Canizares 1997). This method relies on the assumption of hydrostatic
equilibrium in the analysis of X-ray intensity and temperature
profiles, from which the gravitational potential may be deduced. The
requisite data are hard to obtain with current X-ray telescopes, and
the analysis is subject to possible systematic uncertainties if the
hot gas is not in a single phase. Furthermore, many of the X-ray halos
may be associated with the group hosting the elliptical rather than
with the galaxy itself.  It is therefore of interest to develop
alternative probes of the mass distribution in the outer parts of
elliptical galaxies.

If we wish to measure the gravitational potential of an elliptical
galaxy unambiguously using kinematic tracer particles, then we need a
set of test bodies whose orbital structure is simple and
observationally well-constrained.  In this paper, we show that the
large, regular systems of faint shells seen in some elliptical
galaxies (Schweizer 1980, Malin \& Carter 1980) offer just such a
tracer.  These shells are believed to be the remnants of a small
galaxy after a head-on collision with a larger system -- any other
collision geometry does not result in extensive, nested shell systems.
From the geometry of the merger, we know that all the stars in each
shell must be on radial orbits with very similar energies.  As we show
below, this particularly-simple orbital structure means that the
kinematics of the shells provide a useful new probe of the
gravitational potential out to large radii in elliptical galaxies.

\section{Shell Galaxies}

Deep photometric observations reveal that around 10\% of elliptical
galaxies contain large numbers of faint concentric annular shells of
enhanced starlight (Malin \& Carter 1983, Schweizer 1983).  A number
of models have been proposed to explain the origin of these ``shell
galaxies.''  For example, Fabian, Nulsen \& Stewart (1980) have
suggested that shock waves in the intergalactic medium could induce
star formation with the observed shell-like structure, and Thomson
(1991) modelled the shells as a type of spiral density wave excited by
the passage of a companion galaxy.  However, perhaps the most widely
accepted model involves the merging of a small satellite with a large
galaxy.  In this paper we examine the implications of this model for
the kinematics of shells.

\subsection{The basic merger model}

Numerical simulations have shown that when a small galaxy collides
head-on with a larger system, a whole series of shell-like structures
is formed (e.g. Quinn 1984).  The small galaxy is disrupted by
the tidal forces of the encounter, but the orbits that the liberated
stars follow in their new host galaxy closely reflect the original
motion of their progenitor galaxy.  These stars hence find themselves
on nearly radial orbits with similar initial positions and velocities.
All of the stars were injected into the host galaxy at the same
moment, but stars with slightly different energies follow orbits with
slightly different periods, and so as time proceeds the merging stars
slowly spread out along a path through the centre of the host galaxy.

As the stars approach the end of each radial excursion through the
galaxy, they slow to a halt before reversing their path, leading to a
large enhancement in stellar density at these points.  Hence, only
stars that lie close to the extreme ends of their orbits will
contribute significantly to the surface brightness of the galaxy.  At
any given moment, only stars with particular orbital periods will have
completed an integer number of radial excursions and thus find
themselves at this phase of their orbits.  The orbital periods
correspond to particular energies, and so we only see stars with these
quantized orbital energies.  Since stars with different energies have
orbits that take them to different extreme radii, we witness this
phenomenon as a series of shells. 

In summary, each shell is made up of stars from the destroyed galaxy
with equal orbital energies.  The shells at different radii contain
stars with different energies, which therefore have completed
different integer numbers of radial excursions since the merger.
Since the stars in any shell have a very small velocity dispersion
compared to that of their host galaxy, these structures show up very
clearly as sharp edges in the photometry of the system.

\subsection{Numerical simulations of shell kinematics}

\begin{figure*}
\setlength{\unitlength}{1.0in}
\begin{picture}(7,3.5)(0.4,0.5)
\put(0,0){\epsfbox{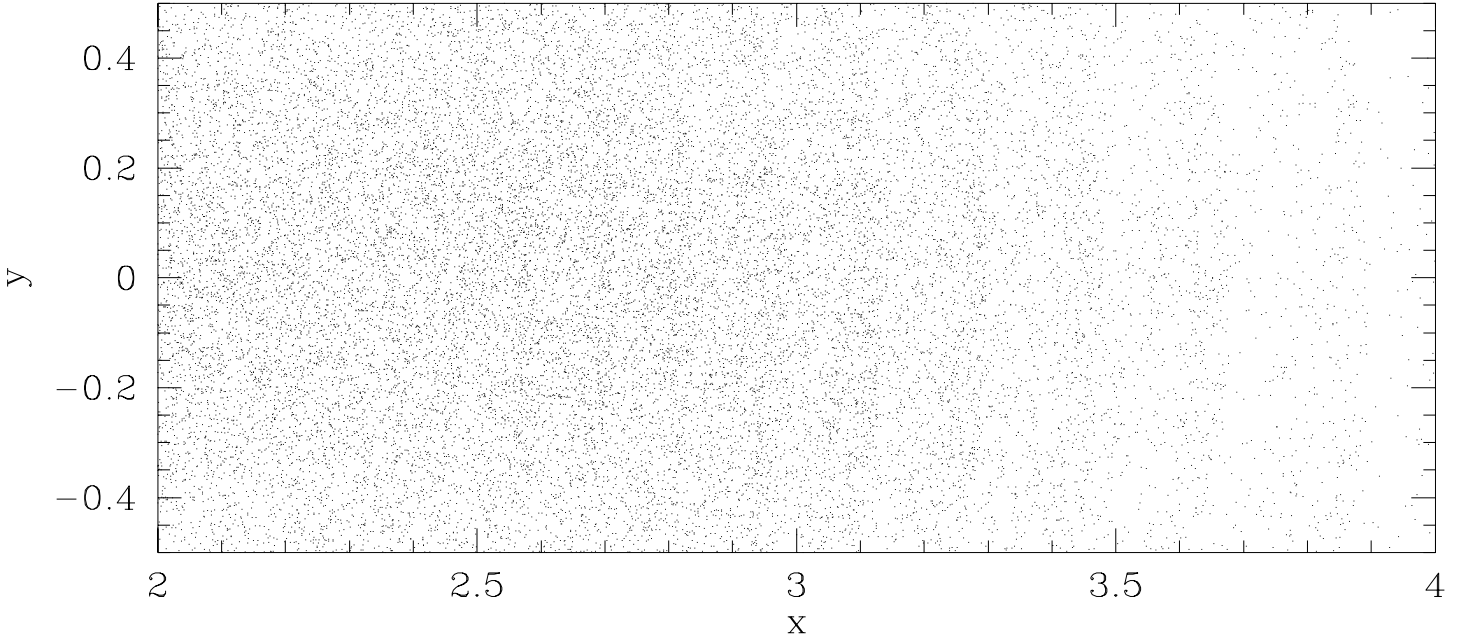}}
\end{picture}
 \caption{Projected density of part of the shell system generated by a
small unbound satellite falling into a spherical isochrone potential.}
\label{fig-shellxy}
\end{figure*}

In order to examine the observable kinematic properties of the shells
produced in a merger, we have carried out simulations in which an
unbound satellite is accreted by a massive spherical galaxy.  We
adopted the isochrone potential,
\begin{equation}
\Psi(r) = -{1\over{1+(1+r^2)^{1/2}}},
\end{equation}
for the host galaxy; since the action/angle variables of the orbits in
such a potential can be calculated analytically (McGill \& Binney
1990; Gerhard \& Saha 1991), the time evolution of the orbits can be
obtained in a single step without requiring full orbit integrations,
and so it is computationally cheap to obtain snap-shot views of the
evolution of such a system containing a large number of test particles.
We modelled the initial configuration for the satellite with a
Gaussian density profile (dispersion 0.5) and zero velocity
dispersion.  Figure~\ref{fig-shellxy} shows the shells produced 750
time units after such a satellite is released with zero velocity at
radius $r=3$.  

\begin{figure*}
\setlength{\unitlength}{1.0in}
\begin{picture}(7,3.5)(0.4,0.5)
\put(0,0){\epsfbox{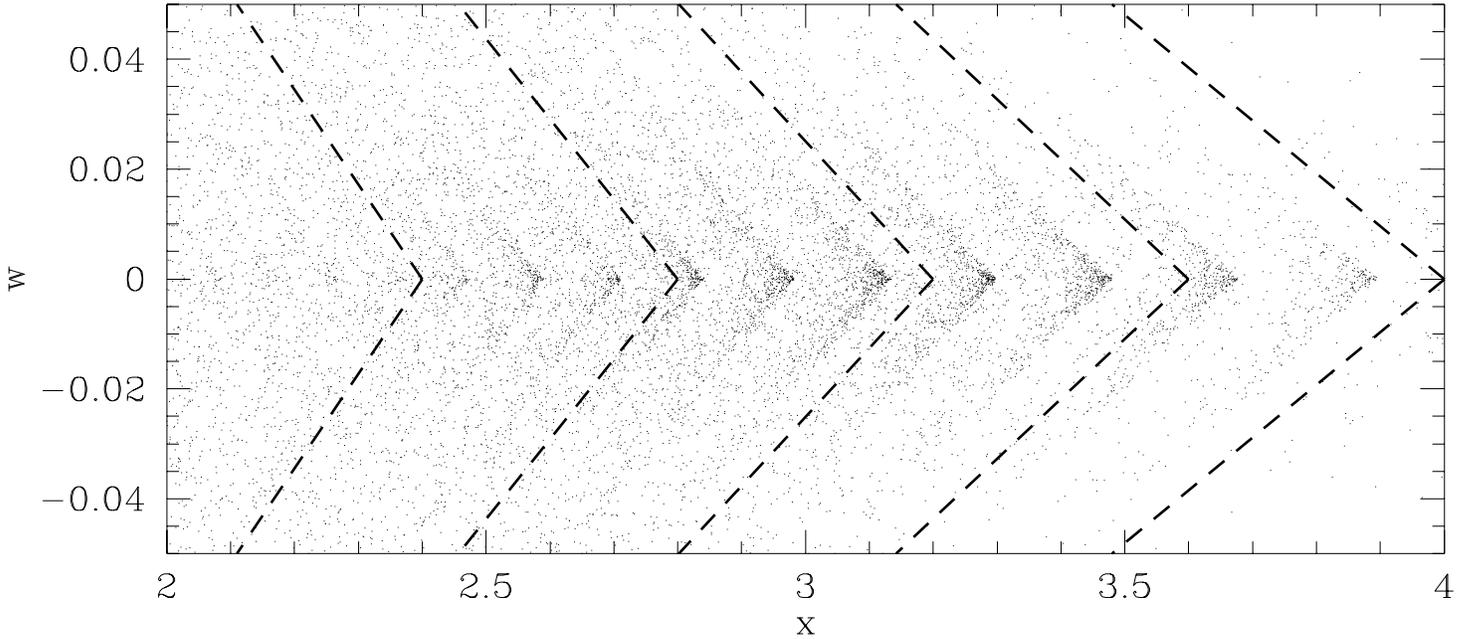}}
\end{picture}
\caption{The velocity structure of the shells shown in Fig.~1. The
slopes of the V-shaped line profiles agree well with the predictions
of the approximate formula of equation~(7) (shown as dashed lines).}
\label{fig-shellxw}
\end{figure*}

Kinematically, what we might hope to observe are the line-of-sight
velocities of the particles in each shell.  Figure~\ref{fig-shellxw}
shows the line-of-sight velocities of the particles as a function of
position along the axis defined by the direction of the collision.
Evidently, each shell has a characteristic projected velocity structure,
with a sharp-edged double-peaked distribution whose width goes to zero
at the shell edge.  Below, we show why such profiles arise, and how
they can be used as diagnostics of the underlying gravitational
potential.
%
%
\def\epsfsize#1#2{1#1}
\def\epsfsize#1#2{1#1}

\subsection{The kinematics of individual shells}

\begin{figure*}
\setlength{\unitlength}{1.0in}
\begin{picture}(7,7)(0.4,0.5)
\put(0,0){\epsfbox{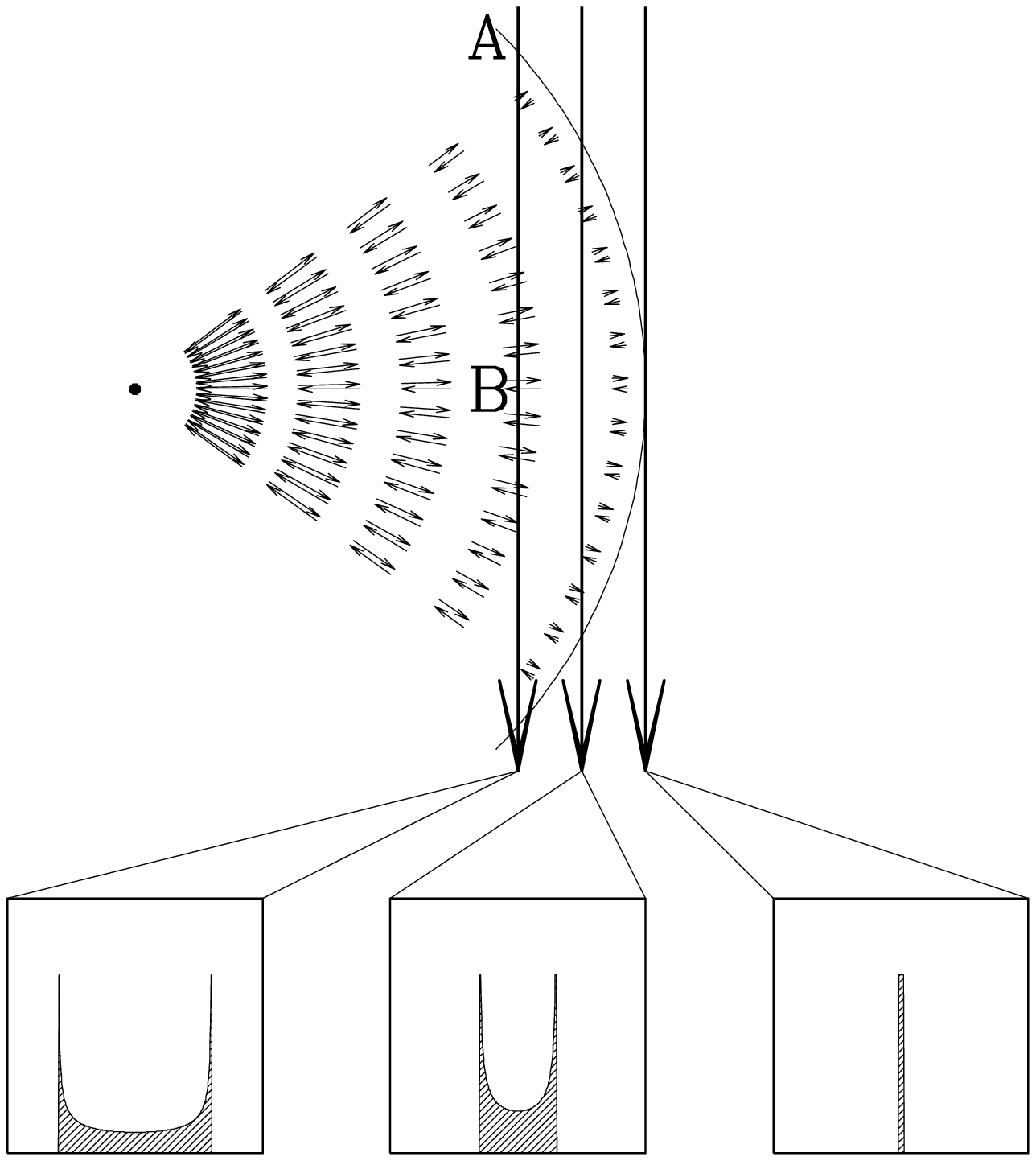}}
\end{picture}
\caption{Schematic diagram showing the velocities of stars in a
spherical shell system. The top part of the diagram shows a
mono-energetic system of stars moving on radial orbits, and the inset
figures show the distributions of these stars' line-of-sight velocities
down three lines of sight.}
\label{fig-scheme}
\end{figure*}

The sharp velocity structures in a shell arises because the shell is
made up of stars with only a small range of orbital energies.  We
might therefore hope to understand Figure~\ref{fig-shellxw} by
treating each shell as a mono-energetic ensemble of stars on radial
orbits.  Figure~\ref{fig-scheme} shows schematically the observable
kinematics -- density of stars as a function of line-of-sight velocity
and position -- for a cross-sectional view through one such shell.
\def\epsfsize#1#2{1#1}
Clearly, the line-of-sight velocity of a star depends on its location
with respect to the shell edge. Near the outer edge of the shell
(point A in Fig.~\ref{fig-scheme}), stars are close to turning around
in their oribits, and so their velocities will lie near the
systemic velocity of the host galaxy.  Similarly, stars near the
tangent point (point B) move mostly transverse to the line of sight,
also resulting in near-systemic observed velocities.  At intermediate
positions, stars do have finite line-of-sight velocities, however;
moreover, on lines of sight further in from the shell edge, orbital
speeds are larger resulting in larger line-of-sight velocities. The
resulting distributions of velocity are sketched schematically in the
lower panels of Fig.~\ref{fig-scheme}.

To derive the variation in maximum velocity with distance from the
shell edge, consider a mono-energetic population of stars orbiting
radially in the gravitational potential $\Psi(r)$. The stars form a shell 
whose outer edge lies at radius $r = r_s$.
By energy conservation, the radial
velocity of stars in this shell at radius $r<r_s$ is simply
\begin{equation}
v_r=\pm \left(2[\Psi(r_s)-\Psi(r)]\right)^{1/2}.
\end{equation}
If $R$ and $z$ denote the projected radius of a line of sight probing
the shell and the distance along this line respectively (with the host 
galaxy at $z=0$), then the
line-of-sight velocity of this material, $\vlos$, is given by
\begin{equation}
\vlos^2=\left({z\over r}v_r\right)^2=2\left(1-{R^2\over r^2}\right)
                    \left[\Psi(r_s)-\Psi(r)\right]
\label{eq-vlossq}
\end{equation}
The maximum value for $\vlos^2$ along any given line of sight can be
calculated approximately by expanding equation~(\ref{eq-vlossq}) about
$r=r_s$:
\begin{eqnarray}
\vlos^2&=&-2(r-r_s)\Psi'(r_s)\left(1-{R^2\over r_s^2}\right)\nonumber\\
       &&-(r-r_s)^2\left({4R^2\Psi'(r_s)+r_s(r_s^2-R^2)\Psi''(r_s)
                     \over r_s^3}\right)                    \nonumber\\
       & &+O\left[(r-r_s)^3\right].
\end{eqnarray}
Along lines of sight close to the shell edge ($|R-r_s|\ll r_s$), this
expression reduces to 
\begin{eqnarray}
\vlos^2&=&4(r-r_s)[\Omega(r_s)]^2(R-r_s)-4(r-r_s)^2\Omega(r_s)^2 \nonumber\\
       &=&4(r_s-r)(r-R)[\Omega(r_s)]^2,
\end{eqnarray}
where $\Omega(r)=[\Psi'(r)/r]^{1/2}$ is the circular frequency at a radius
$r$.

For a given shell radius and line of sight, $|\vlos|$ is maximized at
$r=\half(R+r_s)$, where it has the value
\begin{equation}
v_{\rm max} = \pm\Omega(r_s)(r_s-R).
\end{equation}
In other words, the projected velocity width of a shell at projected
radius $R$, $\Delta \vlos(R)=2|v_{\rm max}|$ increases approximately
linearly inwards from zero at the edge of the shell, with slope
\begin{equation}
{d\Delta\vlos\over dR}=-2\Omega(r_s).
\label{eq-vslope}
\end{equation}
Lines of this slope are plotted in Figure~\ref{fig-shellxw}, and,
indeed, match the extreme velocities of the shell features very well.

Since this slope is just the circular frequency at the radius of the
shell, it provides direct information on the host galaxy's
gravitational potential at this point.  Measuring the slope for each
of the shells then yields a series of measurements of the gradient of
the galaxy's potential at different radii.  These constraints can
either be treated as direct measurements of the enclosed mass at
different radii, or they can be approximately numerically integrated
over radius to obtain a direct measure of the gravitational potential.

In practice, it will be difficult to measure accurately the edges of
the velocity distribution of a shell. What is actually seen in the
spectrum of a shell is the convolution of a stellar spectrum with the
line-of-sight velocity distribution (LOSVD) of the stars in the shell,
$F(\vlos)$.  In order to interpret spectra of shells, we need to know
the shape of this function.  The LOSVD can be calculated analytically
for mono-energetic shells of stars on radial orbits as follows. The
contribution to the distribution by a portion of shell material at
distance $z$ down the line of sight in an area of size $A$ centered on
projected radius $R$ is given by:
\begin{equation}
F(\vlos)d\vlos=\nu(r)Adz,
\end{equation}
where:
\begin{eqnarray}
z^2&=&r^2-R^2;\\
\nu(r)&=&{k\over v_r r^2};\\
v_r&=&[2(\Psi(r_s)-\Psi(r))]^{1/2}; \hbox{\  and}\\
\vlos&=&2\Omega(r_s)[(r-R)(r_s-r)]^{1/2}.
\end{eqnarray}
A given value of $|\vlos|$ corresponds to two distinct radii, 
\begin{equation}
r_\pm=
\half\left\{R+r_s\pm\left[(r_s-R)^2-\Omega(r_s)^{-2}\vlos^2\right]^{1/2}\right\},
\end{equation}
so two contributions need to be added in order to calculate the line
profile:
\begin{eqnarray}
F(\vlos)= {kA|\vlos|\over2\Omega(r_s)^2} 
         \biggl( {1\over r_+z_+v_{r_+}|R+r_s-2r_+|} \hskip 0.5in \nonumber \\
 + {1\over r_-z_-v_{r_-}|R+r_s-2r_-|}\biggr),
\end{eqnarray}
where $z_\pm=(r_\pm^2-R^2)^{1/2}$.
Simplifying by expanding this equation about radius
$r=r_s$, and assuming $(r_s-R)\ll r_s$, we obtain the LOSVD
\begin{equation}
F(\vlos)\propto{1\over\left[(r_s-R)^2-\Omega(r_s)^{-2}\vlos^2\right]^{1/2}}.
\end{equation}
\label{eq-losvd}
This two-horned profile, reminiscent of integrated {\sc Hi} profiles
of disk galaxies, forms a single-parameter family of distributions.
Some sample LOSVDs of this form are illustrated in
Fig.~\ref{fig-scheme}.  By convolving a stellar ``template'' spectrum
with an LOSVD of this form, it will be possible to find the values for
$\Omega(r_s)$ that best reproduce the observed galaxy spectrum.  By
repeating this process for spectra of shells spanning a wide range in
radii, it is thus possible to reconstruct the complete gravitational
potential of the galaxy.

\section{Discussion}

In this paper, we have set out to show how the kinematics of the
shells produced when a small galaxy merges with a large system can be
used to measure the underlying gravitational potential.  As mentioned
earlier in the paper, although the minor merger model is the most
widely accepted interpretation of extensive shell systems, several
other possible explanations have been advanced.  Thus, not only would
the detection of the characteristic kinematic structure illustrated in
Fig.~\ref{fig-shellxw} allow us to measure the gravitational
potential, but it would also confirm the origin of the shells
themselves.  

An alternative method for using shells to constrain the form of the
gravitational potential was originally proposed by Quinn (1984).  He
suggested that use be made of the distribution of shells with radius.
Recall that, at any given time, the photometrically-observed shells
are delineated by those stars that have completed integer number of
radial oscillations.  The ratio of the radii of successive shells
therefore gives the oscillation periods at different energies, which
is directly related to the form of the potential.  The constraints on
the potential obtained from such analyses are non-dimensional; they
could, for example, provide information about the power-law index of
the potential, but they cannot yield a mass normalization.  By
contrast, the kinematic method described here relies on measured
velocities, which can be translated directly into the
fully-dimensional gravitational potential.  The shell counts have also
been shown to be susceptible to uncertainties associated with
dynamical friction acting on the shredding satellite, since different
shells will have followed substantially different orbits, with
different numbers of passages through the center of the host galaxy.
Quinn's method also requires that we know how many radial oscillations
the outermost shell has executed, which can only be inferred
indirectly.  The method discussed here, on the other hand, only uses
information from each shell singly, and only uses the fact that phase
wrapping is an efficient mechanism for concentrating stars of very
nearly equal energies.

Our analysis has assumed that the merger occurred on an exactly radial
orbit. To test the importance of this assumption, we have repeated our
simulations in the isochrone potential for satellite galaxies on
slightly non-radial orbits. These calculations revealed that even in
systems where the encounter is sufficently off-axis to produce shells
that do not appear regular and aligned, the line profiles of
individual shells are still accurately reproduced by the radial model
calculations.  Thus, the kinematics of an aligned shell system will
almost certainly be well-modelled by the simple radial orbit
approximation.  Furthermore, altering the viewing angle of the shell
system tends to diminish the shell visibility before altering the
velocity structure, again suggesting that by selecting aligned,
regular shell systems, we will preferentially find systems to which
the above analysis is applicable.  It should be noted, however, that
the tests we have carried out have all been based on the isochrone
potential, since the orbit calculations are computationally simple in
a potential of this form.  It would be interesting to carry out a more
extensive numerical study of shells in other potentials in order to
check that they, too, are insensitive to the assumption of radial
orbits.

Any flattening of the host galaxy's potential is also a possible
problem.  It may introduce orientation-dependent effects, as well as
modify the orbit structure of the stars making up each shell.
Simulations of the type described earlier in flattened potentials show
that regular shell systems only occur for nearly radial
encounters in spherical potentials: a degree of flattening as small as
5\% in the potential will cause significant beating between the
motions in the various directions, and leads to shells with alternately
large and small opening angles. The velocity structure of such shells
is also rather disturbed, and it is not clear that the velocity
profiles will provide much information about the overall potential.
However, the detection of a regular uniform shell system provides us
with strong {\it a priori} evidence that the potential must be very
close to spherically-symmetric.

Shells are photometrically faint structures, and so obtaining the
kinematic measurements that we advocate here will not be simple.  It
is, however, noteworthy from a comparison of Figs.~\ref{fig-shellxy}
and \ref{fig-shellxw} that the kinematic observation of these systems
significantly enhances their contrast against any background.  It is
also worth noting that the brightnesses of observed shells do not
decrease very rapidly with radius, and so, unlike other techniques for
measuring gravitational potentials using stellar kinematics, the
difficulty of applying this method does not increase prohibitively
with radius.  We have carried out signal-to-noise ratio calculations
for some of the brighter shell galaxies such as NGC~3923, and have
ascertained that data of the requisite quality could be obtained with
a couple of nights' integration using a 4-metre telescope.  Clearly,
projects of this nature will prove quite feasible with the up-coming
generation of large telescopes.

\section*{ACKNOWLEDGEMENTS} 
 
This paper was completed during a visit to the Lorentz Centre at the
University of Leiden, and the authors gratefully acknowledge the
hospitality of this institution.  MRM is supported by a PPARC Advanced
Fellowship (B/94/AF/1840).

\end{document}